%% file: geomatch_rct.tex
\documentclass[12pt]{article}

\usepackage[margin=1in]{geometry}
\usepackage[T1]{fontenc}
\usepackage[utf8]{inputenc}
\usepackage{lmodern}
\usepackage{amsmath,amssymb}
\usepackage{booktabs}
\usepackage{graphicx}
\usepackage{caption}
\usepackage{setspace}
\usepackage[numbers,sort&compress,super]{natbib}
\usepackage{lineno}
\usepackage[hidelinks]{hyperref}
\usepackage{comment}

\newcommand{\dropcap}[1]{#1}

\title{AI-based matching improves refugee employment in a double-blind randomized trial}

\author{Kirk Bansak\textsuperscript{1,2,*}, Jens Hainmueller\textsuperscript{2,3,*,\dag}, Dominik Hangartner\textsuperscript{4,5,*},\\
Jeremy Ferwerda\textsuperscript{2,6}, Elisabeth Paulson\textsuperscript{2,7}, Angie Delevoye\textsuperscript{2},\\
Nicholas Adams-Cohen\textsuperscript{2}, Ashwin Ramaswami\textsuperscript{2}, Selina Kurer\textsuperscript{4,5},\\
Joelle Pianzola\textsuperscript{4}, and Michael Hotard\textsuperscript{2}\\[6pt]
{\small \textsuperscript{*}These authors jointly led this project.}\\
{\small \textsuperscript{\dag}Corresponding author: \href{mailto:jhain@stanford.edu}{jhain@stanford.edu}}}
\date{}

\usepackage{pdfpages}
\hypersetup{pdftitle={AI-based matching improves refugee employment in a double-blind randomized trial},pdfauthor={Bansak, Hainmueller, Hangartner, Ferwerda, Paulson, Delevoye, Adams-Cohen, Ramaswami, Kurer, Pianzola, Hotard}}
\begin{document}

\begin{singlespace}
\maketitle

\vspace{-1.75em}
{\footnotesize
\begin{center}
\textsuperscript{1}Department of Political Science, University of California, Berkeley, Berkeley, CA, USA\\
\textsuperscript{2}Immigration Policy Lab, Stanford University, Stanford, CA, USA\\
\textsuperscript{3}Department of Political Science, Stanford University, Stanford, CA, USA\\
\textsuperscript{4}Immigration Policy Lab, ETH Zurich, Zurich, Switzerland\\
\textsuperscript{5}Public Policy Group, ETH Zurich, Zurich, Switzerland\\
\textsuperscript{6}Department of Government, Dartmouth College, Hanover, NH, USA\\
\textsuperscript{7}Technology and Operations Management Unit, Harvard Business School, Boston, MA, USA
\end{center}}
\vspace{-1em}
{\small
\begin{center}
\textbf{Abstract}
\end{center}
\vspace{-1em}
\noindent
Refugee integration is a central policy challenge for host countries, and
where governments
initially place refugees shapes their integration trajectories. Yet
placement officers often have limited information about where each case is
most likely to succeed. Algorithmic refugee matching uses administrative data,
machine learning, and constrained optimization to recommend
employment-optimized placements in real time as cases arrive, with human
placement officers retaining final authority.
Between January 2020 and June 2023, the Swiss State
Secretariat for Migration randomly assigned about 2,000 refugee cases to
receive a canton recommendation either algorithmically optimized for
employment or drawn to approximate existing procedures, with placement officers and
refugees blinded to assignment.
The two arms used identical but separate canton and
origin-group quotas, so gains reflect better refugee-canton matching rather
than reallocation toward stronger labor markets. The trial began just
before the COVID-19 pandemic shifted labor-market conditions. 
For the pre-registered primary outcome---the share of months employed during the
first three years---the pooled intention-to-treat (ITT) estimate across the
2020--2023 placement cohorts was $+2.2$ percentage points ($\approx 10\%$ of the 22.3\% control mean; 95\% CI [$+0.05$, $+4.33$]), rising
to $+3.9$ pp ($\approx 17\%$; [$+1.11$, $+6.68$]) for the post-COVID 2022--2023 cohorts.
Effects grew over time: at 36 months, the pooled ITT on the employment
rate was $+5.2$ pp ($\approx 11\%$; 95\% CI [$+1.10$, $+9.25$])---comparable to the gains
from hundreds of hours of intensive language training. Overall, the results provide rare field evidence that AI-based decision support can improve high-stakes public-sector allocation,
offering a scalable, low-cost way to raise refugee employment.

\par}
\end{singlespace}
\clearpage

\setcounter{linenumber}{1}

\input{body.tex}

\section*{Acknowledgements}

We thank the Swiss State Secretariat for Migration (SEM) for providing the
administrative data and, above all, for implementing the pilot within the
agency's core placement processes. We are grateful to the SEM placement
officers and staff who carried out the randomized recommendations in their
daily work. We thank Duncan Lawrence for useful comments, George Wood
for research assistance, and Michael Maurer for assistance with the
GeoMatch interface.

\section*{Funding}

The tool development and evaluation were supported by Stanford University's Institute for Human-Centered
Artificial Intelligence, Schmidt Futures, Google.org, Rockefeller
Foundation, Stanford Impact Labs, Charles Koch Foundation, and Open Society
Foundations.
D.H. acknowledges funding from the European Research Council under the European
Union's Horizon 2020 research and innovation programme (grant 804307). The
funders had no role in study design, data collection or analysis, the
decision to publish, or manuscript preparation.

\section*{Author Contributions}

K.B., J.H., and D.H. jointly led the project and are designated as lead authors.
K.B., J.H., D.H., and J.F. conceptualized and designed the study. K.B. and
J.H. developed the core GeoMatch algorithm, and K.B. and E.P. developed the
minimum-discord online assignment rule. K.B., N.A.-C., and A.R.
developed the software interface. K.B., J.H., D.H., J.F., S.K., J.P., and M.H.
coordinated and implemented the trial. K.B., J.H., D.H., A.D., N.A.-C., and
M.H. acquired and curated the data. K.B., J.H., and D.H. conducted the
statistical analysis and prepared the visualizations. K.B., J.H., D.H., J.F.,
E.P., A.D., and M.H. interpreted the results. J.H. and D.H. supervised the
project and acquired funding. J.H. wrote the original
draft. All authors reviewed and revised the manuscript and approved the final
version.

\section*{Competing Interests}

The authors declare no competing interests. The research team received no
payments from SEM or other Swiss government entities for developing or operating
the tool.

\section*{Data Availability}

The sensitive individual-level administrative data are restricted under
the data-use agreement with the Swiss State Secretariat for Migration (SEM).
Researchers may apply directly to SEM, subject to legal, privacy, and
institutional approval.

\section*{Code Availability}

The code used to produce all analyses, tables, and figures will be archived in a public repository upon publication.

\section*{Additional Information}

Supplementary Information is appended to this preprint.
The pre-analysis plan is available on the OSF Registries \citep{pap2023}.
Correspondence should be addressed to Jens
Hainmueller (\href{mailto:jhain@stanford.edu}{jhain@stanford.edu}).

\bibliographystyle{naturemag}
\bibliography{references}

\clearpage
\includepdf[pages=-]{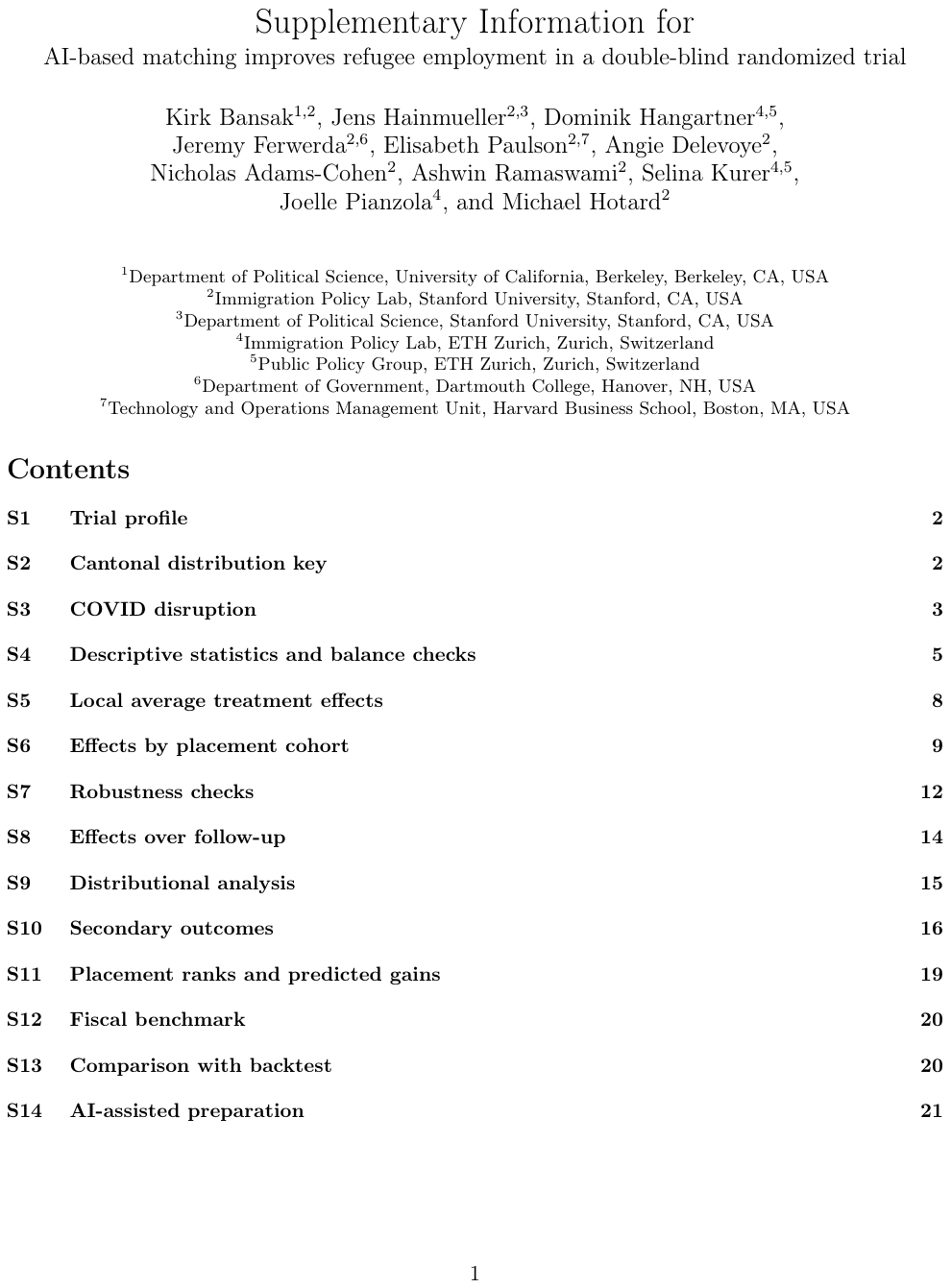}
\end{document}

%% file: body.tex
\dropcap{U}NHCR projects that 136 million people will be forcibly displaced
or stateless worldwide by the end of 2026 \citep{unhcr2026}.
In host countries that grant refugees protection, their labor-market integration is a
central policy challenge. Refugees face many barriers to employment, and
integration is often slow.
One study, for instance, showed that across twenty European countries refugees are about
twelve percent less likely to be employed than
otherwise comparable migrants, a gap that persists for ten to fifteen
years after arrival \citep{fasani2022}. Low employment is costly for
refugees, who depend on limited amounts of public assistance, 
as well as for host societies, which bear welfare costs and forgo
the tax revenue employment would generate.

Research consistently identifies the early period after a 
refugee's arrival in a host country, often called the
``integration window,'' as decisive. Conditions and interventions in the first
months and years after arrival have lasting effects on refugees'
trajectories, making early action more cost-effective than later
remediation \citep{hainmueller2016, marbach2018}. Yet interventions
aimed at this window have yielded inconsistent results \citep{bahar2024, ferwerda2025integration}. Intensive language
training, wage subsidies, and bundled employment programs can improve
employment prospects, but they require substantial investment \citep{arendt2024, marbach2025, dahlberg2024, abbiati2025}. Lighter-touch
programs, on the other hand, often have limited or no lasting effects \citep{emeriau2025, battisti2019}. Cost-effective and scalable interventions remain elusive.

A separate stream of research highlights the importance of where refugees are settled. In many destination countries, refugees are assigned to regions or
municipalities according to administrative distribution rules, such as
population-based quotas, rather than based on where they are most likely to succeed.
A large body of quasi-experimental evidence shows that these initial placement decisions can have large and persistent effects on refugees'
employment and earnings \citep{aslundrooth2007, dammrosholm2010, azlor2020, godoy2017, auer2018, schmid2023, marten2019}.
Yet existing allocation systems rarely leverage these place effects, because
placement officers typically have limited information about where each refugee
would fare best while also needing to satisfy complex distribution requirements.

Adding algorithmic recommendations to the placement process offers the
potential to capture these gains within existing admissions policies, quotas,
and service constraints. First introduced by Bansak et al.
\citep{bansak2018}, outcome-based algorithmic refugee matching combines machine learning
with constrained optimization to achieve precisely that goal. Under this framework, predictive models estimate how each refugee
would fare across possible locations, and an optimization step generates
recommended matches between refugees and available slots to maximize predicted
aggregate employment while respecting capacity limits and other assignment
rules. Holding geographic quotas fixed, the approach does not direct more
refugees toward stronger labor markets; instead, it helps placement officers
identify better-fitting locations for individual refugees. Because it relies
on routinely collected administrative data and operates within the existing
placement process, it has low marginal costs. The approach is also advisory by
design in that it solves a complex placement problem that officers could not practically evaluate
case by case, recommends a location, and leaves officers with final authority to
incorporate information not captured in the administrative data.

Backtests using historical placement 
data suggest potential employment gains, and tools based on this approach have been
piloted in Switzerland, the Netherlands, the United
States, and Canada \citep{bansak2026robustness}. Subsequent research has
extended these methods with online assignment and dynamic
resource balancing, welfare and fairness
constraints, and learning under distribution shift and across
small locations \citep{ahani2021placement,acharya2020matching,freund2023group,ahani2024dynamic,bansak2024outcome,bansak2024learning,jain2025ctrl,bansak2026dynamic}. 
Yet every published estimate of the approach’s benefits rests on retrospective backtests, which cannot fully account for implementation frictions, staff responses, noncompliance, general-equilibrium displacement, or changing conditions between training and deployment. Whether these estimated gains carry over to real-world implementation therefore remains a central empirical question.

This study reports, to our knowledge, the first randomized controlled trial of algorithmic refugee matching, conducted in partnership with the Swiss State Secretariat for Migration (SEM), 
the federal authority responsible for Switzerland's asylum system. 
The SEM was the first government agency
to adopt these methods in the refugee placement setting,
deploying and testing an early generation of a tool named GeoMatch. More broadly, the trial provides rare field evidence on whether AI decision-support systems built on predictive models can improve high-stakes public-sector decisions.

Between January 2020 and June 2023, the SEM randomly assigned eligible
refugee cases to receive either an algorithmic recommendation (treatment arm) or a
status quo recommendation (control arm) with equal probability.
The algorithmic recommendation selected the canton predicted to maximize
aggregate employment across cases, subject to strict national origin
distribution quotas across cantons.
The status quo recommendation randomly selected a canton without using
employment predictions, subject to an identical but separate copy of the
quotas, so that the two arms never competed for the same allocation slots.
Both placement officers and refugees were blinded to treatment status.
The algorithm was trained and trial outcomes were analyzed using SEM
administrative register data linked to employment histories. Treatment effects are estimated at the level of the refugee case, which
may comprise a family or an individual and is the unit of randomization
and placement.
The primary outcome is the mean share of months employed among adult case
members during the first three years following placement in their assigned
canton. To
characterize how effects evolve within this window, the analysis also examines
point-in-time employment across the follow-up period, including whether any adult in
the case is employed 36 months after placement. Consistent with the target in the pre-analysis plan (PAP)
\citep{pap2023}, the analysis sample includes the first
2{,}000 cases with complete three year employment outcomes in order of
assignment; results are nearly identical when all 2{,}046 complete cases
through June 2023 are included (Supplementary Information, Sec.~S7). 

Attrition was low and balanced: 2.5\% of control-arm
adults and 2.3\% of treatment-arm adults were no longer observed at three years (Supplementary
Information, Sec.~S4).
Baseline covariates were balanced across treatment arms
(\(F_{11,1{,}999}=0.49\), \(p=0.91\)). The mirrored quota design also produced
balance across arms in both canton placements and the joint distribution of
canton and origin group (\(\chi^2\) tests \(p>0.99\); Supplementary Information,
Table~S3).
Thus, in this trial, the treatment contrast isolates gains from improved
refugee-canton matching while holding the aggregate canton and origin-group
allocation fixed.

The primary analysis reports intention-to-treat (ITT) effects of assignment to receive a \mbox{GeoMatch} recommendation rather than a status quo recommendation. It also considers local average treatment effects (LATE) for placement in the GeoMatch-recommended canton, taking into account noncompliance with recommendations. The analyses use pre-registered covariate-adjusted
ordinary least squares (OLS) and instrumental variables (IV) regression
specifications (see Methods). Alternative estimators and adjustment sets yield nearly
identical results (Figure~1).
Secondary and diagnostic analyses examine alternative employment measures, 
secondary migration, placement ranks, and distributional effects.

The trial provides a conservative test, for reasons anticipated in the PAP and
detailed in the Methods.
First, the prediction stage of the algorithm used a sparse predictor set drawn
from the government's administrative register and selected by the SEM for
deployment.
The Swiss register data did not include standard human capital variables
that are known to be predictive of employment and are usually available
in other contexts, such as education, occupation, prior work experience,
and host-language proficiency.
Second, to preserve implementation fidelity and keep the treatment definition
fixed, the model specification and optimization procedure were not changed as
newer matching methods became available.
Third, the placement process was tightly constrained by Switzerland's cantonal
proportionality key and by origin-group balancing rules.
These constraints limited the feasible assignment set and therefore the scope
for the algorithm to exploit location-specific synergies, especially those
involving ethnic networks or language region matches
\citep{marten2019, auer2018, schmid2023}.
In addition, the mirrored quota design meant that the algorithm optimized
over only about half of the caseload and its slots, further shrinking the
feasible match set relative to a full-scale deployment.
Finally, the trial launched in January 2020, just before the COVID-19 pandemic
disrupted cross-border mobility, asylum processing, and labor markets. This created a gap
between the data available for model training and the conditions in which the tool was used. Prior scholarship has documented how COVID-19 disruptions led to abrupt distribution shifts and thereby created challenges for prediction and forecasting across medical, economic, business, and other domains \citep{kagerbauer2024susceptibility,andonov2023impact,yin2022covid,papik2023impacts,castle2021value,o2022business}. These disruptions likely made it harder for models trained on pre-pandemic data to predict employment for the trial's early cohorts.

Despite these challenges, the trial shows positive pooled gains of roughly
10\%, with larger gains of roughly 17--21\% for the post-COVID
cohorts whose integration windows unfolded after
the labor market had mostly normalized---a pattern consistent with the
algorithm's predictions realigning with realized
outcomes after the pandemic.
Treatment effects also grow over follow-up. At 36 months after
placement, when refugee employment becomes economically meaningful,
algorithmically guided placement raised the share of cases with any
adult employed by roughly 11--12\%---an effect comparable to the gains
from hundreds of hours of intensive language training (see Discussion).

\section*{Results}\label{results}

\subsection*{Main Treatment Effects}\label{main-treatment-effects}

Table~1 reports the ITT and LATE estimates for the
pre-registered primary outcome---the share of months employed over the
first three years after placement---aggregated at the case level as the
mean across adult case members. Pooled across all placement cohorts, the case-level ITT estimate is \(+2.2\) percentage points (pp) on a control mean of \(22.3\%\) of months employed, a relative increase of \(\approx 10\%\), with a 95\% confidence
interval from \(+0.05\) to \(+4.33\) pp and \(p=0.045\) (\(N=2{,}000\) cases; Table~1). The case-level LATE is \(+2.5\) pp (\(\approx 11\%\), 95\% CI \([+0.05,+4.93]\) pp, \(p=0.046\)).
The LATE first stage is strong: randomization increased placement in the
GeoMatch-recommended canton by approximately 88 pp (partial
\(F=7{,}107\); Supplementary Information, Table~S5).

\begin{table*}[t!]
\centering
\caption{\textbf{Treatment Effects on Three Year Employment.}}
\small
\resizebox{\textwidth}{!}{%
\begin{tabular}{llcccccc}
\toprule
Sample & Estimand & $N$ & Ctrl. mean (\%) & $\beta$ (pp) & Gain & 95\% CI (pp) & $p$ \\
\midrule
All cohorts (2020--2023) & ITT & 2{,}000 & 22.3 & $+2.19$ & $+9.8\%$ & $[+0.05,+4.33]$ & 0.045 \\
                    & LATE & 2{,}000 & 22.3 & $+2.49$ & $+11.1\%$ & $[+0.05,+4.93]$ & 0.046 \\
\midrule
Excluding peak COVID (2021--2023) & ITT & 1{,}709 & 22.0 & $+2.71$ & $+12.3\%$ & $[+0.39,+5.03]$ & 0.022 \\
                    & LATE & 1{,}709 & 22.0 & $+3.11$ & $+14.1\%$ & $[+0.45,+5.77]$ & 0.022 \\
\midrule
Post-COVID (2022--2023) & ITT & 1{,}212 & 22.3 & $+3.90$ & $+17.5\%$ & $[+1.11,+6.68]$ & 0.006 \\
                    & LATE & 1{,}212 & 22.3 & $+4.59$ & $+20.6\%$ & $[+1.35,+7.82]$ & 0.006 \\
\bottomrule
\end{tabular}
}
\begin{flushleft}
\footnotesize \textit{Notes:} Estimates report ITT and LATE effects.
The outcome is the mean share of months employed over the
first three years after placement among adult case members. ITT estimates capture access to a GeoMatch algorithmic recommendation. LATE estimates capture initial placement in the GeoMatch-recommended canton
among compliers, instrumenting placement with randomization. Control means are shown as percentages, and the effects in the column labeled
$\beta$ are in percentage points.
The column labeled Gain denotes the effect in terms of a percentage increase over
the baseline, which is calculated by dividing the percentage-point effect
by the corresponding control mean. The ``Excluding peak COVID'' row drops the 2020 launch cohort but retains the
2021 transition cohort and the post-COVID cohorts.
The ``Post-COVID'' row pools the 2022--2023 post-COVID cohorts.
\(N\) counts the number of cases.

\end{flushleft}
\end{table*}

The pooled estimates may, however, mask cohort heterogeneity
associated with the COVID-19 disruption. The analysis therefore reports
non-prespecified comparisons across placement cohorts. GeoMatch was initially trained on pre-pandemic
administrative data, and the PAP warned that COVID-era labor markets
might not reflect the relationships learned from those data
\citep{pap2023}. The 2020 launch cohort began its integration window during the acute
pandemic disruption, and the 2021 cohort remained heavily exposed during
early integration. By contrast, the 2022--2023 cohorts mostly integrated
after the disruption had passed and COVID-related restrictions had been
lifted (Supplementary Information, Sec.~S3). Table~1 accordingly reports estimates excluding the peak-COVID 2020
cohort and estimates restricted to the post-COVID 2022--2023 cohorts.

Estimates are larger once the peak-COVID launch cohort is excluded: the
case-level ITT rises to \(+2.7\) pp (\(\approx 12\%\); 95\% CI
\([+0.4,+5.0]\) pp, \(p=0.022\); \(N=1{,}709\) cases) and the LATE to
\(+3.1\) pp (\(\approx 14\%\); 95\% CI \([+0.4,+5.8]\) pp,
\(p=0.022\)). In the post-COVID 2022--2023 cohorts, the ITT reaches
\(+3.9\) pp (\(\approx 17\%\); 95\% CI \([+1.1,+6.7]\) pp,
\(p=0.006\); \(N=1{,}212\) cases) and the LATE \(+4.6\) pp
(\(\approx 21\%\); 95\% CI \([+1.4,+7.8]\) pp, \(p=0.006\)).

In sum, algorithmic recommendations raised employment on the pre-registered
primary outcome, with gains roughly twice as large once labor markets had
stabilized. A direct comparison across eras supports this pattern. Specifically, the
estimated treatment effect is \(+4.6\) pp larger in the post-COVID
2022--2023 cohorts than in the COVID-era 2020--2021 cohorts
(treatment-by-era interaction \(p=0.039\) with covariate adjustment, \(p=0.070\) unadjusted).
Cohort-by-cohort estimates show the same progression---effects statistically 
indistinguishable from zero in 2020 and 2021, and then effects that are
positive, statistically significant, and
stable across the 2022 and 2023 cohorts (Supplementary Information,
Sec.~S6, Figs.~S4 and~S6; Fig.~S5 shows the same progression using
double/debiased machine learning).

\subsection*{Robustness of the Main Estimates}\label{robustness}

Figure~1 shows that the estimates are stable across alternative
specifications. The PAP-primary specification adjusts for covariates by interacting
treatment with mean-centered covariates, following Lin \citep{lin2013}
(see Methods). The alternatives include the unadjusted and
post-double-selection estimators \cite{belloni2014high},
inverse-probability-weighted attrition correction, the same
specification without canton indicators, and an additive
specification. For the LATE, the analysis considers three definitions of treatment take-up: initial placement in the GeoMatch-recommended canton, placement plus at least 18 months of residence in the recommended canton, and the share of follow-up spent in the recommended canton. 
For all three analysis samples, the
resulting ITT and LATE estimates are virtually unchanged across specifications. The full set of estimates
is reported in the Supplementary Information (Tables~S5 and~S6). Additional robustness checks show that the individual-level estimates are
close to the case-level estimates,
with similar magnitudes, precision, and cohort patterns (Supplementary Information,
Table~S7). 

\begin{figure*}[t!]
\centering
\caption{\textbf{Robustness of the Three Year Employment Effects.}}
\includegraphics[width=\textwidth]{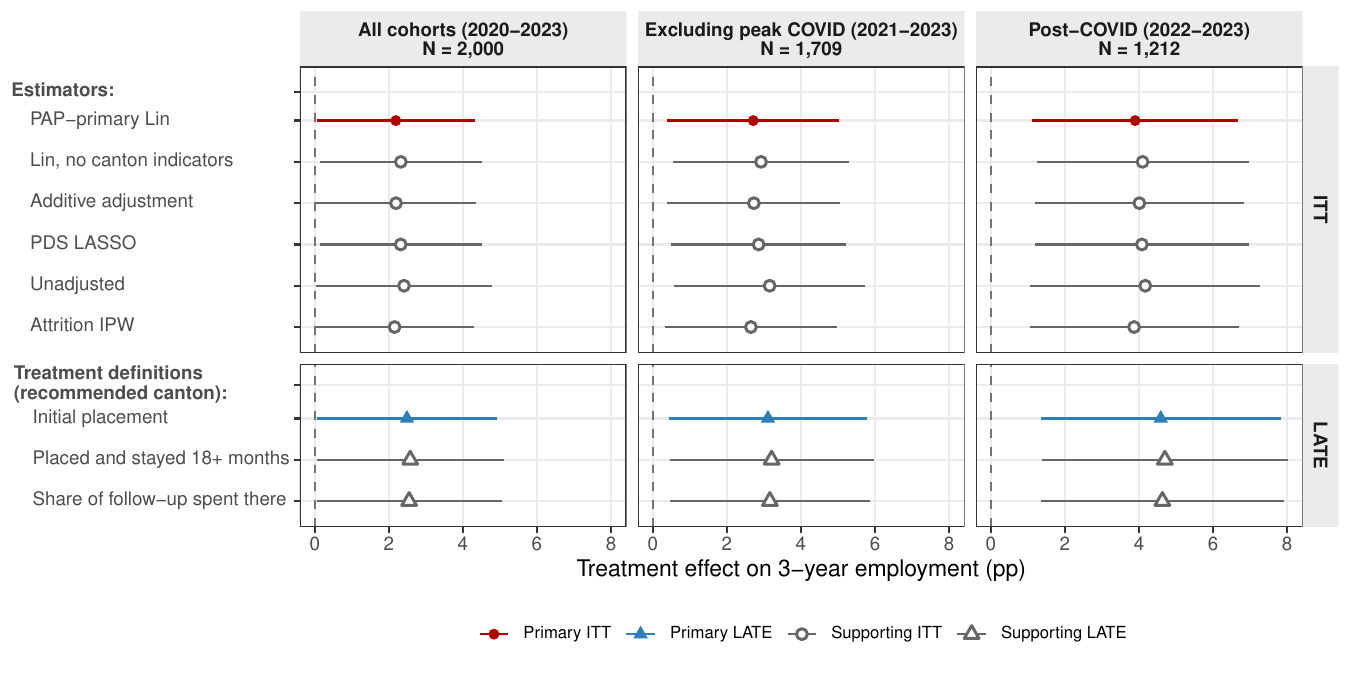}
\begin{flushleft}
\footnotesize \textit{Notes:} The top panel shows ITT
estimates for three year employment under alternative estimators. The
bottom panel shows LATE estimates under the alternative
definitions of treatment take-up given in the text.
Red and blue points mark the primary
ITT and LATE specifications; grey points show supporting analyses.
PDS denotes post-double-selection, and IPW denotes inverse-probability
weighting. Columns correspond to the three samples in
Table~1. Points show estimates and bars show 95\% confidence intervals.
Full estimates are reported in
the Supplementary Information, Tables~S5 and~S6.
\end{flushleft}
\end{figure*}

\subsection*{Effect Growth over Follow-up}\label{effect-growth}

\begin{figure*}[t!]
\centering
\caption{\textbf{Employment Effects Grow Over Follow-up.}}
\includegraphics[width=\textwidth]{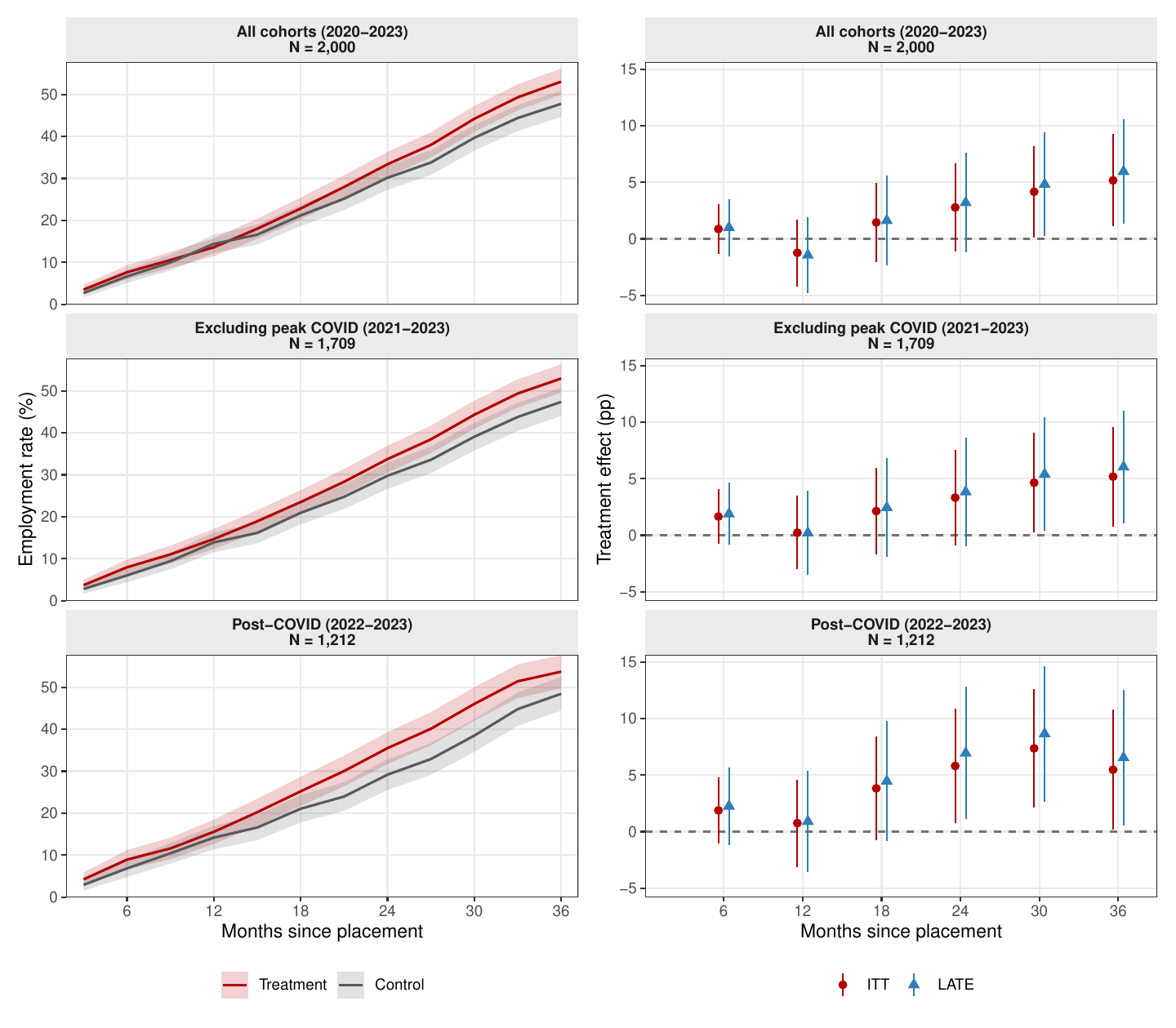}
\begin{flushleft}
\footnotesize \textit{Notes:} Rows correspond to the three samples
evaluated in Table~1. Left panels show the case employment rate, by assignment arm, at quarterly horizons from 3 to 36 months after placement; shaded bands are pointwise 95\% intervals. Right panels
report the corresponding case-level ITT and LATE
estimates at six-month intervals. Bars show 95\% confidence intervals. Panel headers report the 36-month sample sizes. The 36-month estimates
are reported in the Supplementary Information,
Table~S9.
\end{flushleft}
\label{fig:growth-main}
\end{figure*}

Refugee labor-market integration typically progresses slowly in Switzerland
and elsewhere in Europe \citep{hainmueller2016,marbach2018,fasani2022}, making
it important to examine whether the effects of improved placement grow with
time since placement.
Figure~\ref{fig:growth-main} traces the evolution of employment over
follow-up for the same cohort samples as Table~1. The left panels show the
employment rate---the share of cases with any adult employed---at
quarterly intervals since placement, separately by assignment arm. This
point-in-time measure was not specified in the PAP but is built from the
same monthly employment indicators as the primary outcome. It also
corresponds more closely to how employment is conventionally measured in
labor economics.

Employment rises steadily from low
levels in both arms, the two trajectories track each other closely through
the first year, and the treatment arm then pulls ahead, with the gap
widening through the end of follow-up. The right panels report the
corresponding case-level ITT and LATE estimates at six-month intervals.
Across samples, point estimates become larger
at longer horizons. In the pooled sample, the ITT estimate is close to zero
at early horizons but grows to $+5.2$ pp at 36 months (95\% CI $[+1.1,+9.2]$; $\approx 11\%$ of the 47.7\% control share), with the
corresponding LATE at $+6.0$ pp (95\% CI $[+1.4,+10.6]$) and the post-COVID
ITT reaching $+5.5$ pp ($[+0.2,+10.8]$). A complementary analysis of the share of months
employed over the first one, two, and three years shows the same pattern of
increasing effects at both the case and individual levels (Supplementary Information,
Fig.~S7).
Overall, the estimated effects grow over follow-up rather than
fading, and the pooled estimates are largest at the longest observed
horizon.

\clearpage

\subsection*{Distributional Analysis}\label{distributional-analysis}

The Supplementary Information presents empirical cumulative distribution functions
(ECDFs) of the primary three year employment outcome by
assignment arm for the three samples considered in Table~1 (Fig.~S8). Across all three samples, the treatment ECDF lies below the control ECDF over
almost the entire support, with the largest separation in the post-COVID
sample. Two complementary distributional tests formally assess this pattern
(Supplementary Information, Table~S10). First, a test for first-order stochastic dominance (FOSD), based on the Barrett--Donald
supremum statistic, asks whether the treatment outcome distribution weakly
dominates the control distribution, meaning that the treatment ECDF is nowhere
above the control ECDF \citep{barrett_donald_2003}. The test fails to reject in
all three samples (all $p\geq0.81$), indicating that the data are statistically
consistent with treatment FOSD over control. Second, a one-sided
Kolmogorov--Smirnov test rejects distributional equality in favor of a
treatment-favoring shift in the pooled sample ($D^+=5.41$ pp, $p=0.035$), the
sample excluding 2020 ($D^+=6.47$ pp, $p=0.018$), and the post-COVID sample
($D^+=8.76$ pp, $p=0.007$). Overall, the increasingly pronounced distributional shift
mirrors the pattern in the mean effects. From a social planning perspective, the
broad rightward shift indicates that the treatment regime improves the
distribution of employment outcomes across its range rather than in a
narrow segment.
Likewise, a refugee choosing behind a veil of ignorance about their own
circumstances would have reason to prefer the treatment regime's employment
prospects.

\subsection*{Secondary Outcomes}\label{secondary-outcomes}

\begin{figure*}[t!]
\centering
\caption{\textbf{Effects on Secondary Employment Outcomes.}}
\includegraphics[width=0.96\textwidth]{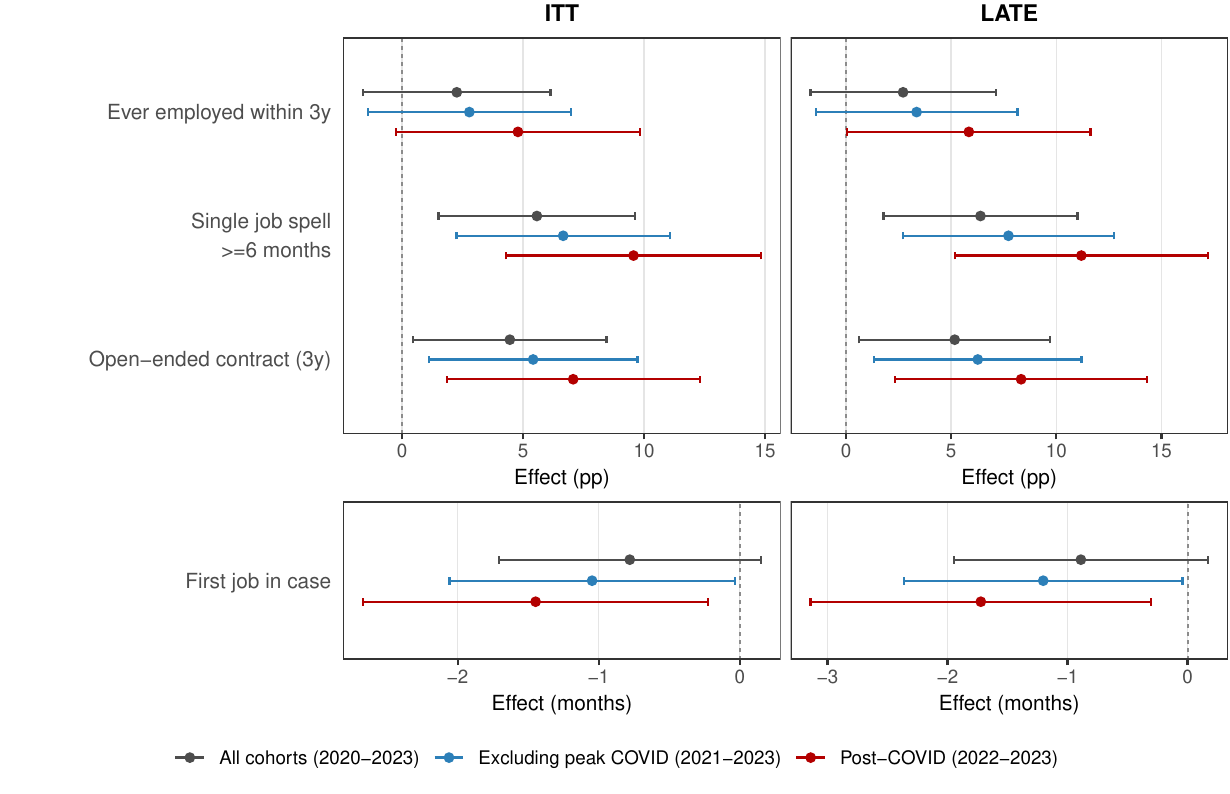}
\begin{flushleft}
\footnotesize \textit{Notes:} Points show
covariate-adjusted case-level ITT and LATE estimates for four outcomes: whether any adult in
the case worked at any point during
the first 36 months, whether any adult had a single job spell covering at least 183 days
(six months), whether any adult had an open-ended contract at three years,
and months until the first
adult in the case obtained employment, censored at 36 months. Negative estimates
for time to first employment indicate faster employment. Bars show 95\% confidence
intervals. Outcome definitions and additional results are reported in
Supplementary Information, Sec.~S10 and Table~S11.
\end{flushleft}
\label{fig:secondary-employment-stability}
\end{figure*}

Figure~\ref{fig:secondary-employment-stability} reports results using secondary
employment outcomes. The prespecified
any-employment outcome indicates whether any adult in the case worked at any point
during the first three years, even briefly. Additional outcomes not specified in the PAP capture
the speed and stability of employment. Any-employment estimates are positive
but fairly imprecise, with a pooled ITT of \(+2.26\) pp (95\% CI \([-1.61,+6.13]\) pp, \(p=0.253\)) that rises to \(+4.78\) pp in the post-COVID cohorts (95\% CI \([-0.25,+9.81]\) pp, \(p=0.063\)).

The evidence is stronger for sustained employment. In the pooled sample,
assignment to the treatment arm raises the probability that any adult has a
single job spell of at least six months by \(5.57\) pp in the ITT analysis
(95\% CI \([1.50,9.64]\) pp, \(p=0.007\)). The ITT rises to \(9.56\) pp
in the post-COVID cohorts (95\% CI \([4.30,14.82]\) pp, \(p<0.001\)).
Treatment also raises the probability that any adult has an open-ended
contract at three years by \(4.45\) pp in the pooled sample
(95\% CI \([0.46,8.44]\) pp, \(p=0.029\)) and \(7.07\) pp in the
post-COVID cohorts (95\% CI \([1.84,12.29]\) pp, \(p=0.008\)).
The first adult obtains employment an estimated \(0.78\) months sooner in
the pooled sample (95\% CI \([-0.15,1.71]\) months, \(p=0.100\)) and
\(1.45\) months sooner in the post-COVID cohorts
(95\% CI \([0.22,2.67]\) months, \(p=0.021\)). Taken together, these results
suggest that access to algorithmic placement recommendations shifted cases
toward earlier and more sustained employment, particularly in the later
cohorts.

Secondary migration out of the initial placement canton within three years is
rare in the control arm (5.4\% of cases) and not differentially affected by
treatment (\(+0.35\) pp, \(p=0.71\); Supplementary Information, Table~S11), consistent with
legal restrictions on cantonal mobility.

\clearpage

\subsection*{Prediction Ranks and Realized Gains}\label{prediction-fit}

\begin{figure*}[t!]
\centering
\caption{\textbf{Placement Ranks, Prediction Fit, and Realized Gains.}}
\includegraphics[width=0.90\textwidth]{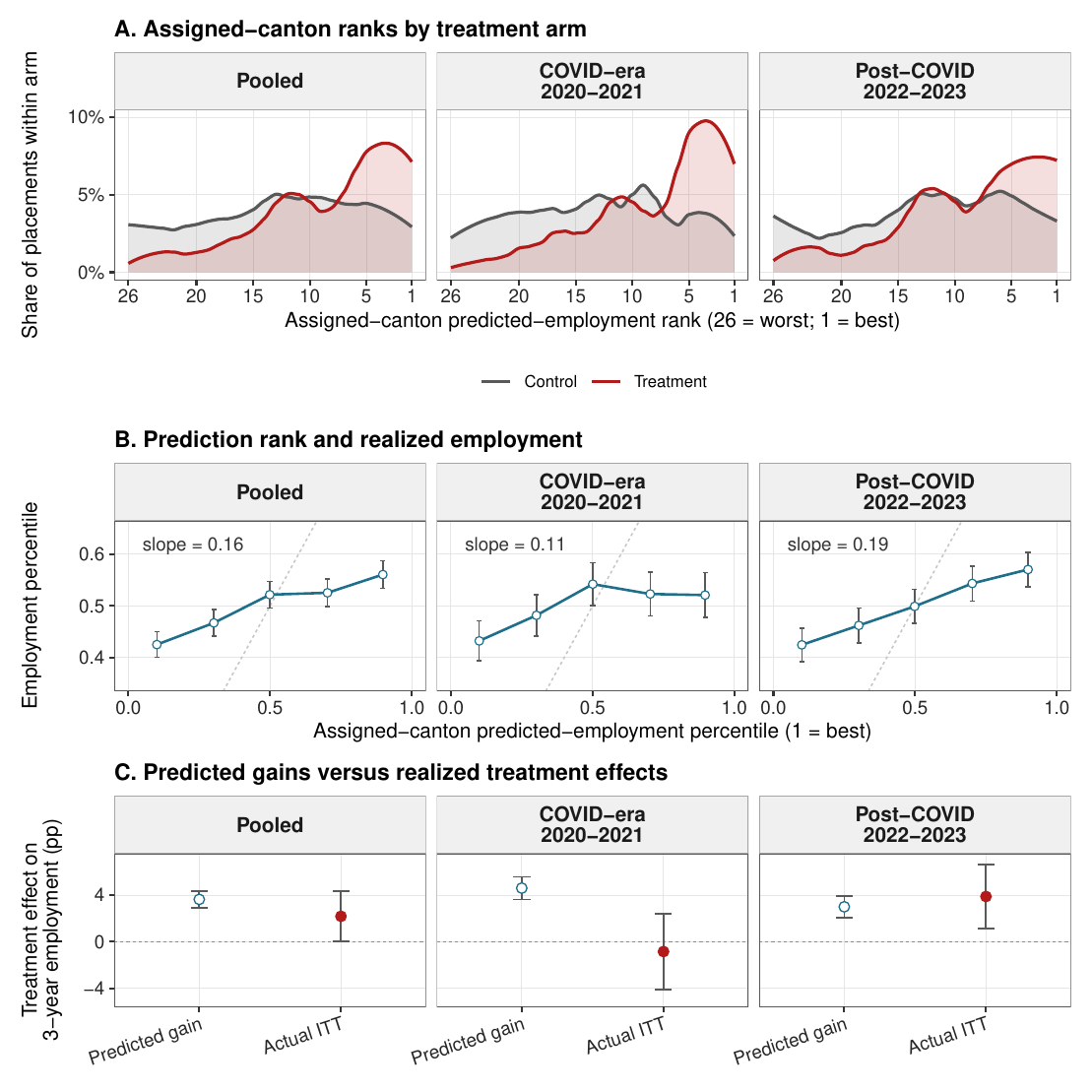}
\begin{flushleft}
\footnotesize \textit{Notes:} Columns report pooled cohorts, COVID-era cohorts
(2020--2021), and post-COVID cohorts (2022--2023).
Panel A plots the arm-specific distribution of assigned-canton predicted-employment ranks
\((1=\) highest predicted employment, \(26=\) lowest).
Panel B relates the predicted-employment percentile of the assigned canton to
the realized employment percentile.
Panels A and B use adult-level prediction profiles and orient the horizontal
axis so worse-ranked cantons are on the left and better-ranked cantons are on
the right.
Panel C aggregates to the case level, comparing the predicted employment gain
at realized placements with the case-level ITT.
Intervals are 95\% confidence intervals where applicable.
Construction details for all three panels are provided in the Supplementary Information,
Sec.~S11.
\end{flushleft}
\label{fig:prediction-mechanism}
\end{figure*}

To better understand why the estimated gains are larger in later cohorts,
Figure~\ref{fig:prediction-mechanism} examines two links required for
algorithmic matching to improve employment. First, recommendations must
shift placements toward cantons ranked more highly within each refugee's
predicted-employment profile. This occurred in both the COVID-era and
post-COVID cohorts: treatment cases were disproportionately placed in
cantons near the top of their own rankings. Second, higher assigned-canton
ranks must be associated with better realized outcomes. During the COVID
era, the rank-on-rank relationship is essentially flat in the upper half,
where most treatment placements occur, indicating little ability to
distinguish among the cantons most often recommended. In the post-COVID
cohorts, by contrast, the relationship is stronger and increases
monotonically. Consistent with these patterns, the model predicts a
positive treatment-control employment difference in both periods, while
the realized ITT gain is substantially larger in the post-COVID cohorts.
Taken together, the diagnostics show that treatment shifted placements
toward higher-ranked cantons throughout the trial, but that these
rankings translated into larger realized gains after the pandemic. This
pattern is consistent with COVID-induced distribution shift: labor-market
conditions during the early trial diverged from the algorithm's
pre-pandemic training data, weakening the alignment between predicted
rankings and realized employment. As conditions stabilized, that
alignment strengthened and treatment gains increased.

\section*{Discussion}\label{discussion}

This is, to our knowledge, the first randomized controlled trial of
algorithmic refugee matching, an approach that has attracted
interest from governments and resettlement agencies because of its potential to
improve employment at low marginal cost. Even when averaged across a deployment period
spanning the COVID-19 disruption, the pooled estimates imply employment gains of
roughly 10--11\%. Among the post-COVID cohorts, the
estimated gains rise to roughly 17--21\% across the ITT and LATE.
The effects also grow over the three year follow-up, consistent with the
gradual pace of refugee labor-market integration. At 36 months, treatment
raises the probability that any adult in the case is employed by $+5.2$
to $+6.0$ pp ($\approx 11$--$12\%$ across the ITT and LATE), a
particularly policy-relevant margin. In sum, the results indicate that algorithmically guided placement
raised refugee employment on the order of 10--20\% in a real-world
deployment. 

Several features make the trial a
demanding and deliberately constrained test. First, the trial spanned the
COVID-19 disruption, and the COVID-era cohorts weigh down the pooled
estimates. Indeed, the post-COVID cohorts show gains roughly twice as large.
Second, the trial evaluated a tightly constrained deployment, with a
sparse predictor set, a low-base-rate outcome measured over a long horizon,
stringent balancing constraints that precluded many potentially beneficial
matches, and mirrored quotas that limited the optimizer to about half of
the caseload and its slots. The estimates therefore capture a restricted version
of the approach. The Methods section describes these issues in greater detail,
and the Supplementary Information compares the trial estimates with the original
retrospective backtest predictions (Sec.~S13).

Cost-benefit calculations, which use the fiscal benchmark detailed in the
Supplementary Information, Sec.~S12, suggest that these employment gains are
large relative to implementation costs.
The central benchmark values each additional employment-month at CHF
1,950 (all monetary amounts are in Swiss francs; CHF 1 \(\approx\) USD
1.2 at September 2026 exchange rates), combining reduced public support payments with social security
contributions and source taxes.
Implementation costs are assumed at CHF 50 per treated adult, covering the
additional operating expenditure of layering the tool onto the existing
placement workflow. This allowance includes placement-officer time, computing,
software maintenance, and routine monitoring. One-time development and
integration costs are excluded from this figure because they amortize at
scale.
On this basis, the individual-level pooled ITT estimate of about 1.95 percentage points---equivalent to 0.70 additional employment-months per adult over three years---translates into roughly CHF 1,300 in fiscal benefits per adult net of implementation costs. The resulting benefit-cost ratio, calculated using gross fiscal benefits, is approximately 27:1.
At a scale of 1,000 treated adults, the pooled ITT corresponds to about CHF 1.30 million in net fiscal benefits.
These calculations are deliberately narrow, in that they capture
only fiscal returns to the government---reduced support payments plus the
payroll contributions and source taxes collected on refugees' earnings---and
do not count the earnings that refugees retain or other welfare gains that
accrue to refugees themselves. Even under this narrow accounting, the estimated
fiscal benefits substantially exceed implementation costs.

These gains are comparable to those of substantially more
resource-intensive refugee labor-market interventions, but are achieved
at far lower marginal cost.
In the most comparable metric, the trial's individual-level ITT on being
employed three years after placement is \(+4.6\) percentage points (LATE \(+5.4\) pp) on a control mean of 44.8\% (Supplementary Information, Table~S9).
Denmark's 1999 reform, which added roughly 430 hours of language training,
increased full-time-equivalent employment by 4.2 percentage points on average
over 18 years \citep{arendt2024}, while Germany's 600-hour integration
courses increased
employment by roughly 5 percentage points after one year \citep{marbach2025}. Randomized trials of
labor-intensive programs---including a Swedish package combining language
instruction, work practice, and job search assistance \citep{dahlberg2024}, as well as an Italian program
providing job mentoring and subsidized internships \citep{abbiati2025}---reported gains of 15 and 10
percentage points, respectively. These programs cost roughly CHF 2{,}400 and CHF 3{,}000
per participant and require substantial staffing and delivery capacity, making
them resource-intensive to scale. Meanwhile, lighter-touch interventions often yield
limited or no lasting effects
\citep{emeriau2025,battisti2019}. To date, no other causally evaluated refugee-integration intervention has
combined employment gains of the order observed in the trial with such low marginal cost. This
cost advantage arises because algorithmic matching improves a placement
decision the government makes anyway rather than adding a costly new program.
This comparison should not be interpreted as evidence that algorithmic matching
can substitute for intensive language courses.

Beyond refugee placement, the trial provides rare randomized field
evidence on prediction-based AI decision support in public
administration, complementing randomized evidence on statistical
support systems for caseworkers in employment offices
\citep{behncke2009targeting} and on algorithmic risk assessments
provided to judges in criminal justice \citep{imai2023algorithm}. Its timing also illustrates the vulnerability of such
systems to distribution shift: the largest gains occur among cohorts
that integrated after labor-market conditions normalized, and prediction
diagnostics show closer alignment between predicted and realized gains
in those cohorts. These findings underscore the importance of monitoring
predictive fit, detecting distribution shift, and retraining as
conditions change \citep{bansak2024learning,jain2025ctrl}. More broadly,
they show that predictive tools can improve consequential public
decisions \citep{kleinberg2015prediction} when used to support, rather
than replace, human decision-makers and when operating within explicit
policy constraints.

Future randomized evaluations should test upgraded systems in other
institutional settings and under different allocation constraints.
Priorities include richer predictor sets, shorter-horizon proxy outcomes
that permit faster updating, dynamic and multi-objective optimization,
and methods designed explicitly for distribution shift. That the highly
constrained implementation increased employment during a period of
substantial disruption and at low marginal cost suggests that updated
and less constrained systems could generate even larger gains.

\section*{Methods}\label{materials-and-methods}

\subsection*{Trial Eligibility}\label{trial-eligibility}

The cases in the trial entered the SEM's normal asylum process as asylum seekers.
Eligible cases were assigned by the SEM to the fast-track procedure
(\textit{beschleunigtes Verfahren}) and, before cantonal placement, had
obtained either subsidiary protection (permit F) or recognized refugee status
(permit B). They were also ``free'' cases not subject to legally required
placement in a specific canton and had at least one adult member.

\subsection*{Algorithmic Recommendation System}\label{a-deliberately-frozen-early-generation-algorithm}

The intervention evaluated in this trial was an early generation version of
GeoMatch, an outcome-based refugee-matching system based on the methodology in
Bansak et al. \citep{bansak2018}.
The system recommended one of Switzerland's 26 cantons for each eligible case.
A case is the placement unit, consisting of one or more refugees (typically a household or
family) who are placed together.
For each canton, a separate stochastic gradient-boosted tree model predicted a
working age refugee's three year employment proportion from a small set of
baseline characteristics selected by the SEM for deployment: gender, age at
arrival, marital status, case size, arrival year and month, nationality, and
native language.

The models were trained on administrative data from past arrivals and tuned by
cross-validation \citep{bansak2018,pap2023}.
During the trial, the models were retrained roughly quarterly using the
newest available administrative data. This scheduled retraining is part of the GeoMatch design,
through which the deployed system learns from accumulating data. The predictor
set, model class, optimization target, and constraint structure remained
fixed.
The optimization target, chosen by the SEM, was the cumulative share of months
employed over the first three years after placement.

Because each case is placed as a single unit, the refugee-level predictions
within a case were combined into one case-level metric.
For the employment-share outcome optimized in the trial, the
tool followed the SEM's requested aggregation rule and averaged the adult
members' predicted employment proportions within each case.
A constrained optimization procedure then recommended a canton for each case to
maximize aggregate predicted employment, subject to the system's
capacity constraints.

The SEM imposed two stringent optimization constraints, reflecting rules that
govern its regular cantonal allocation process \citep{asylv1}.
First, every placement had to respect Switzerland's cantonal proportionality
key, which allocates asylum cases across the 26 cantons roughly in proportion
to canton population.
Second, the key is crossed with nationality quotas requiring several
national-origin groups to remain separately balanced across cantons in
the same proportions \citep{pap2023}.
In addition, placements were made online and case by case: cases arrived sequentially and had
to be placed before later arrivals were known.
To handle this online problem, the early generation system used a
minimum-discord allocation rule \citep{bansak2024outcome}.
The rule simulates the expected future caseload and recommends the feasible
canton that least disrupts the allocation of slots anticipated for future
arrivals, thereby preserving scarce slots for later cases likely to need
them while respecting capacity constraints.

GeoMatch ran as a standalone, human-in-the-loop decision support tool built
with input from placement officers to fit their existing workflow. For each eligible case, the
placement officers received a recommended canton that could be accepted or
overridden, retaining final placement authority.

\subsection*{Experimental Design}\label{the-trial}

The trial was authorized by the SEM Board and designed in consultation
with the cantons represented in the tripartite working group AGNA
({\it Arbeitsgruppe zur Neustrukturierung des Asylbereichs}).
Before the trial began, it was reviewed and approved by the Stanford University
Institutional Review Board (protocol 46216), and the ETH Zurich Ethics
Commission accepted the Stanford approval as sufficient for ETH-based
investigators.
The Committee for the Protection of Human Subjects at Dartmouth College also
approved the study (STUDY00031234).
The Stanford University Institutional Review Board waived the requirement for
informed consent because the intervention was embedded in routine
administrative placement and the analysis used administrative records.
The trial was a single-intervention, double-blind randomized controlled trial with one
treatment arm and one control arm.
The PAP specified a target of 2,000 cases, with 1,000 cases per arm, and the
case was the unit of randomization.
This sample-size target was agreed with the SEM rather than derived
from a formal power calculation, and it was fixed before any outcome
data were received.

Each eligible case was randomized independently, with probability 0.5, to
receive either an
algorithmic recommendation (treatment) or a status quo recommendation
(control).
Placement officers screened cases for trial eligibility and entered them into
the placement tool, where assignment was executed by a computer-generated
random draw. Allocation was therefore determined centrally and
automatically after eligibility had been established: a case's
assignment did not exist at the time of screening and was never revealed
afterward, ruling out selective enrollment based on expected treatment
status. The tool's interface
was identical in both arms; only the recommended canton differed.
Eligibility criteria, the randomization procedure, and the mirrored
quota design remained unchanged from the January 2020 launch through
June 30, 2023.

The treatment recommendation selected the canton predicted to maximize
aggregate employment across cases, subject to the
quota and balancing constraints described above. The control recommendation was designed to approximate the status quo procedure for free cases. Under regular Swiss dispersal policy, the SEM assigns asylum cases to cantons
according to the cantonal proportionality key and national-origin balancing
rules, using records from ZEMIS (the SEM's Central Migration Information
System) and without personal interaction with applicants.
Refugee preferences are not considered except for narrow legal exceptions,
such as close family ties, health needs, or the accommodation of unaccompanied
minors.
The control arm therefore approximated the status quo for free cases by
randomly selecting among feasible canton slots under the same quota structure, without using employment predictions.

Two design features protected the integrity of the treatment contrast.
First, the design prevented interference through separate, mirrored quotas.
The trial quota was split into identical treatment and control quotas, so each
case drew only on slots reserved for its own arm.
This ruled out interference through slot competition and kept the aggregate
canton and origin-group allocation the same across arms.
As a result, any treatment effect cannot come from assigning more treated cases
to stronger labor markets, but only from changing which cases are recommended
for a particular canton within the same distribution of slots.

Second, the trial was double-blind. For each trial case,
placement officers received either the GeoMatch recommendation or the
control recommendation in the same format and workflow, but were not told
which procedure generated the recommendation.
Refugees were likewise unaware of treatment status.
Thus, officers could respond to the displayed recommendation but not to
its source. Outcomes came from
independently collected register records and were not subject to
assessor discretion based on treatment status.

\subsection*{Pre-Analysis Plan}\label{pre-analysis-plan}

The trial's PAP was registered on the OSF Registries on March 9, 2023,
after the trial launched in January 2020 but before linked employment
outcome data were received \citep{pap2023}.
The PAP defines the primary outcome as the three year employment proportion,
the employment measure optimized by the tool.
The primary estimand is the intention-to-treat (ITT) effect of assignment to
the treatment arm rather than the control arm.
The primary estimator is ordinary least squares (OLS) regression with the
prespecified covariate adjustment from the Lin specification
\citep{lin2013} that interacts the demeaned covariates with the treatment indicator.

The PAP also specifies local average treatment effect (LATE) analyses for
placement in the GeoMatch-recommended canton, allowing for noncompliance with recommendations.
For each case, the GeoMatch-recommended canton is defined as the canton the
algorithm recommended in the treatment arm or would have recommended had the
case been assigned to treatment.
An indicator then records whether the case was initially placed in that
GeoMatch-recommended canton.
The LATE instruments this placement indicator with random assignment to the
treatment arm. Noncompliance arises when officers do not follow the displayed
recommendation (``nonadherence'') or, more commonly, when the control and GeoMatch procedures
recommend the same canton. Officer adherence to the
displayed recommendation was 97\%. Section~S5 of the Supplementary Information details these channels and
the identifying assumptions.
The PAP specifies two additional pre-registered LATE definitions:
initial placement in the GeoMatch-recommended canton combined with
residence for at least half of the three year follow-up, and the fraction
of the follow-up period spent in that canton.
Because secondary migration is rare, these residence-based
definitions produce very similar estimates.

The PAP also pre-registers secondary outcomes and analyses, including any
employment during the three year follow-up, emigration, secondary migration,
and alternative estimators.
The PAP anticipated that the COVID-19 disruption could weaken the performance
of models trained on pre-pandemic data. Analyses by placement cohort and
follow-up horizon provide a natural way to examine this concern, although they
were not explicitly prespecified.

\subsection*{A Conservative Test}\label{why-this-is-a-hard-test}

Several features of the Swiss deployment make the trial a conservative test of
algorithmic refugee matching, as the PAP
anticipated.\footnote{The PAP described the recommendations as a ``highly
constrained version of the GeoMatch algorithm'' whose observed gains would
be a ``lower bound for gains that may be realized in less constrained
versions''; conversely, ``if no gains are observed, this does not imply
that no gains could be realized with less constrained versions''
\citep{pap2023}.}

The trial evaluated an early generation version of the matching system.
Since launch, the methodology has advanced, including richer
predictive models, online assignment with dynamic resource balancing, methods
for learning under distribution shift, and multi-objective trade-off methods
\citep{acharya2020matching,freund2023group,bansak2024outcome,bansak2024learning,jain2025ctrl}.
To preserve implementation fidelity and keep the treatment definition fixed,
the deployed model specification and optimization procedure remained unchanged
as these methods became available. Periodic retraining updated the fitted models
with newly available data but did not alter the underlying method.
The information available to the deployed algorithm was also sparse. 
The ZEMIS register does not currently record standard human capital variables
that are known to be predictive of employment, such as education, occupational skills,
prior work history, and host-language proficiency. %
Predictive accuracy, and therefore the scope for finding person-place synergies,
was bounded by the information the algorithm was allowed to use.
The three year employment outcome itself was difficult to optimize as well, as refugee
employment in this window is low, with a control-arm mean of about 22\% of months employed over three years.
Small relative movements are therefore hard to detect.

The placement geography was also coarse.
Recommendations were made across Switzerland's 26 cantons rather than across
municipalities or local labor markets, and the proportionality key's uneven
distribution left several small cantons with less than 1\% of placements
(Supplementary Information, Table~S1).

The quota constraints described in the experimental design further restricted
the optimization problem.
The cantonal proportionality key was crossed with nationality-balance rules \citep{asylv1}. Thus,
the algorithm had to recommend cases within a fixed canton-by-origin-group
distribution rather than shifting more members of a given origin group to
cantons where that group tends to fare better, for example because of
co-ethnic networks or language region matches.
This constraint cuts against a well-documented finding in the
refugee integration literature that networks and language matches matter for
employment.
The importance of these constraints is demonstrated by prior research in Switzerland, which showed that assignment to a canton with a
10\% larger co-national network raises the probability of employment three
years after arrival by about 0.3 percentage points (roughly 2\% relative)
\citep{marten2019}, while placement on the linguistically matching side of the
German--French language border more than doubles employment over the first
five years after arrival \citep{auer2018, schmid2023}.
The trial therefore measures the value of matching within a heavily constrained
quota structure, not the gains possible under less constrained allocation.

The mirrored quota design imposed one further restriction on the optimizer.
To keep the arms free of interference, the available canton-by-origin-group
capacity was divided into separate treatment and control pools, so the
algorithm optimized assignments over only about half of the eligible
caseload and its slots---a restriction no practical deployment would impose.
Because matching gains grow with the size of the joint pool, particularly
where small cantons hold only a handful of slots, a system-wide deployment
optimizing the full caseload would face a richer set of feasible matches
than the trial allowed.

Two further features added noise and distribution shift.
First, the tool applied only to trial-eligible cases intended to be placeable
in any canton.
In the historical data used for training, however, the available free-case
indicator only imperfectly approximated this eligibility rule.
This mismatch between the training indicator and the trial population
likely diluted performance.
Second, the trial began in January 2020, just before the start of the 
COVID-19 pandemic.
The Swiss Federal Council declared an ``extraordinary situation'' on
March 16, 2020, closing shops, restaurants, and leisure venues, and lifted
many measures only in February 2022 \citep{admin_covid}.

The pandemic shock affected labor markets and the relationship between
employment and geographic placement through several channels
\citep{admin_covid,sem_covid_asylum_2020,kof_labour_challenging_2021,
adecco_job_index_q3_2021}. Border controls and entry restrictions disrupted
mobility. The SEM temporarily modified asylum
interviews and procedural rules, and working hours were cut through
short-time work schemes that prioritized protecting incumbent jobs over
hiring new labor-market entrants. 
Vacancy data show that job advertisements regained
pre-pandemic levels only in the third quarter of 2021, with hospitality
hiring especially depressed until the summer reopening. The asylum employment spell records provide corresponding evidence: among employed cases in
the 2019--2023 arrival cohorts, the main jobs were concentrated in hospitality
(restaurants and hotels) and other face-to-face service work, with sizable
shares also in cleaning and construction-related work.
These entry jobs were plausibly among those most exposed to closures, capacity
restrictions, reduced customer demand, and short-time work during the early
pandemic.

Since integration is front-loaded, the first one to two years after placement
are when refugees typically enter
language and integration programs, build local networks, search for first
jobs, and establish early work histories \citep{hainmueller2016,marbach2018}.
Cohorts placed in 2020 and 2021 therefore experienced pandemic disruptions
during the period when placement quality was most likely to matter.
By contrast, the 2022 and 2023 cohorts began this early integration window after the
most severe mortality waves and broadest labor-market and institutional
disruptions had largely passed (Supplementary Information, Fig.~S2).

The models available to the early trial cohorts were trained on
pre-pandemic data. They were therefore extrapolating person-place employment relationships
from an economy that no longer closely resembled the one those cohorts
entered.
The system also could not learn from COVID-era three year employment outcomes
in real time, because those outcomes were observed only years after placement
and the deployed algorithm did not use short-horizon proxy outcomes for
updating.
Later cohorts integrated under conditions closer to those in the training data.

\subsection*{Data and Sample}\label{data-and-design}

The data come from the SEM's ZEMIS administrative register, which tracks
asylum applicants from arrival through the asylum process. Its
case-processing and placement records are linked to basic demographic
characteristics and employer-reported employment histories. The three year
window is the 36 calendar months following the month of canton assignment; the
partial assignment month is excluded. A month counts as employed if it falls
within an employer-reported employment spell recorded in ZEMIS. Following the
PAP sample-size target, the analytic sample consists of the first
2{,}000 eligible cases with complete three year employment outcomes,
ordered by the date and time of assignment. The source extract covers
assignments between January 2020 and June 30, 2023 and contains 2{,}096
eligible cases comprising 2{,}256 adults present at initial assignment
(Supplementary Information, Fig.~S1). Adults are case members whose age
at arrival is at least 18.
Of these cases, 2{,}046 have complete three year outcomes. Retaining the
first 2{,}000 (1{,}001 treatment and 999 control cases) yields an
analytic sample of 2{,}156 adults, with the final selected case assigned
on June 13, 2023. A sensitivity analysis including
all 2{,}046 complete cases yields nearly identical results (Supplementary
Information, Sec.~S7, Table~S8).

Attrition and inverse-probability-weighting diagnostics use a broader
recruitment frame containing all eligible cases randomized through the
June 13 selection boundary, regardless of three year outcome
completeness. This frame contains 2{,}050 cases comprising 2{,}209
adults. Employment is observed only while an individual has an active
administrative record; individuals no longer active at three years are
likely to have left Switzerland.

In the analytic sample, mean age at arrival is 27.9 years; 16\% are
female; and the largest origin groups are Afghanistan (54\%), Turkey (26\%),
and Syria (9\%).
The mean arrival-to-placement gap is 2.8 months (capped at its 95th
percentile).
In the control arm, the share of months employed is 7\% in the first year, 14\% over the first two years, and 22\% over the first three years.
Among control-arm refugees, 57\% hold any job at some point within the first
three years.

\subsection*{Design Checks}\label{design-checks}

Placement officer adherence to the displayed recommendation was high and
similar across arms. Specifically, eventual placement matched the displayed canton in
roughly 97\% of cases in each arm (97.3\% in control and
96.5\% in treatment; difference \(=-0.8\) pp, \(p=0.37\)).

At the case level, the mirrored quotas produced nearly identical canton
distributions across treatment and
control arms ($\chi^2$ test of independence
\(p>0.99\) in the PAP analysis sample). The same balance holds within
each placement cohort (2020--2023;
$\chi^2$ test \(p > 0.92\) for all cohorts). The joint distribution of canton
and origin group (Afghanistan, Turkey, and Other) was likewise balanced
($\chi^2$ test \(p>0.99\); Supplementary Information, Table~S3 and Fig.~S3).

In the pooled sample, baseline covariate means are nearly identical across arms
(Supplementary Information, Table~S2), and an omnibus regression of treatment
assignment on the listed baseline
covariates does not reject joint balance (\(F_{11,1{,}999}=0.49\),
\(p=0.91\)). Finally, three year outcome observability is balanced. 
The percentage unobserved at three years is 2.5\% in control and 2.3\% in
treatment, with an adjusted treatment-control difference of \(-0.27\) pp
(\(p=0.67\); Supplementary Information, Sec.~S4, Table~S4).

\subsection*{Analysis}\label{analysis}

The ITT captures deployment of the GeoMatch recommendation with the observed
level of officer take-up, including any overrides. The LATE captures the effect
of placement in the GeoMatch-recommended canton among compliers
\citep{angrist1996identification}.

The main analysis uses the refugee case as the unit of analysis,
matching the unit of randomization and placement and the objective
optimized by the algorithm: the mean three year
employment proportion among adult members of the placed case.
The individual-level analog is reported in the Supplementary Information
(Table~S7) and yields similar results.
The ITT estimator is an OLS regression of this case-level outcome on treatment
assignment and covariates, entered with interactions via the Lin specification 
\citep{lin2013}, as registered in the PAP. An instrumental variables
(IV) specification with additive covariate adjustment estimates the LATE. All specifications use robust standard errors clustered by case.
Unless otherwise noted, reported confidence intervals and $p$-values are
two-sided.
The covariate set follows the PAP's primary specification and includes year- and month-of-placement fixed effects, age-bin indicators, gender,
case-size indicators (top-coded at four or more members), marital-status indicators, nationality group indicators
(top three nationalities plus other), language group indicators (top three
languages plus other), time from arrival or entry to placement, and canton fixed
effects.
Case-level covariates use the earliest-arriving adult; when multiple adults tie
on the earliest arrival date, their covariate values are averaged. Additional specifications---including the unadjusted difference in means, the primary Lin specification omitting canton fixed
effects, additive covariate adjustment, 
post-double-selection LASSO \cite{belloni2014high}, and inverse-probability-weighted
attrition correction---are reported in the Supplementary Information and
yield results that are virtually identical to the primary Lin specification
(Table~S6).
Details of AI-assisted manuscript preparation are provided in the Supplementary Information (Sec.~S14).